\documentstyle[12pt,twoside,fleqn,espcrc1,psfig]{article}

\newcommand{ \eps }{{\varepsilon}}

\newcommand{\AmS}{{\protect\the\textfont2
  A\kern-.1667em\lower.5ex\hbox{M}\kern-.125emS}}

\title{Centrality Dependence of Directed and Elliptic Flow at the SPS}

\author{A.M.~Poskanzer$^\textrm{\footnotesize{a}}$ and
S.A.~Voloshin\address{Lawrence Berkeley National Laboratory, 
Berkeley, CA 94720, USA} for the NA49 Collaboration}

\begin{document}
\maketitle

\noindent
J.~B\"{a}chler$^{6}$, D.~Barna$^{5}$, L.S.~Barnby$^{3}$,
J.~Bartke$^{7}$, R.A.~Barton$^{3}$, L.~Betev$^{14}$,
H.~Bia{\l}\-kowska$^{17}$, A.~Billmeier$^{11}$, C.~Blume$^{8}$,
C.O.~Blyth$^{3}$
% ...

\vspace{0.5cm}
\noindent
% ...
$^{3}$Birmingham University, Birmingham, England.\\
% ...
$^{5}$KFKI Research Institute for Particle and Nuclear Physics, Budapest, Hungary.\\
$^{6}$CERN, Geneva, Switzerland.\\
$^{7}$Institute of Nuclear Physics, Cracow, Poland.\\
$^{8}$Gesellschaft f\"{u}r Schwerionenforschung (GSI), Darmstadt, Germany.\\
% ...
$^{11}$Fachbereich Physik der Universit\"{a}t, Frankfurt, Germany.\\
% ...
$^{14}$University of California at Los Angeles, Los Angeles, USA.\\
% ...
$^{17}$Institute for Nuclear Studies, Warsaw, Poland.\\
% ...

\begin{abstract}
New data with a minimum bias trigger for 158 GeV/nucleon Pb + Pb have
been analyzed. Directed and elliptic flow as a function of rapidity of
the particles and centrality of the collision are presented. The
centrality dependence of the ratio of elliptic flow to the initial
space elliptic anisotropy is compared to models.
\end{abstract}

\section{Motivation}
In the Fourier decomposition of the azimuthal distribution of
particles, the first and second coefficients correspond to the
directed, $v_1$, and elliptic, $v_2$, flow,
respectively~\cite{methods}. The elliptic flow is expected to be
sensitive to the system evolution at the time of maximum
compression~\cite{sorge97}. Ollitrault~\cite{olli92} showed that in a
hydro model the elliptic flow is proportional to the initial space
elliptic anisotropy of the overlapping region weighted by the number
of nucleon collisions in the beam direction. This initial space
elliptic anisotropy, which we will call $\eps$, has been calculated
for a Woods-Saxon density distribution and shown to be almost
insensitive to the nucleon-nucleon cross section~\cite{jacobs99}. It
is enlightening to plot $v_2/\eps$ versus
centrality~\cite{sorge99,heisel99} in order to look for changes in the
reaction mechanism or properties of the nuclear matter. Thus the
motivation for this work is to find a signature (elliptic flow), scan
this signature as a function of a control parameter (centrality), and,
after first dividing out the geometry of the initial state, look for a
change in the physics (unexpected behavior).

\section{Experiment}
NA49 has published directed and elliptic flow results from the NA49
Main Time Projection Chambers for a set of data taken with a medium
impact parameter trigger~\cite{PRL,QM97}. We now have a new set of
data taken with a minimum bias trigger so that we can study the flow
centrality dependence. Also, the tracks from the Main and Vertex TPCs
are combined resulting in full coverage of the forward hemisphere. The
data in the graphs below presenting flow as a function of rapidity
have been reflected about mid-rapidity. The data have been integrated
over $p_t$ and in some cases also over $y$ using as weights the
measured double differential cross
sections~\cite{cooper99,sikler99}. The data have been sorted into six
centrality bins using the Zero Degree Calorimeter, with ``cen1'' being
the most central and ``cen6'' the most peripheral. The impact
parameter values for these bins have been estimated from the number of
participants which were obtained by integrating the
yields~\cite{cooper99,sikler99}. Slightly higher values of $b$, used
in the oral presentation of this paper, were determined from the
fraction of the total cross section corresponding to each bin. Only
some of the available data has been analyzed so far. Thus these data
are preliminary and no systematic errors have been included yet.

\section{Results}
\begin{figure}[htb]
\begin{minipage}[t]{80mm}
\psfig{figure=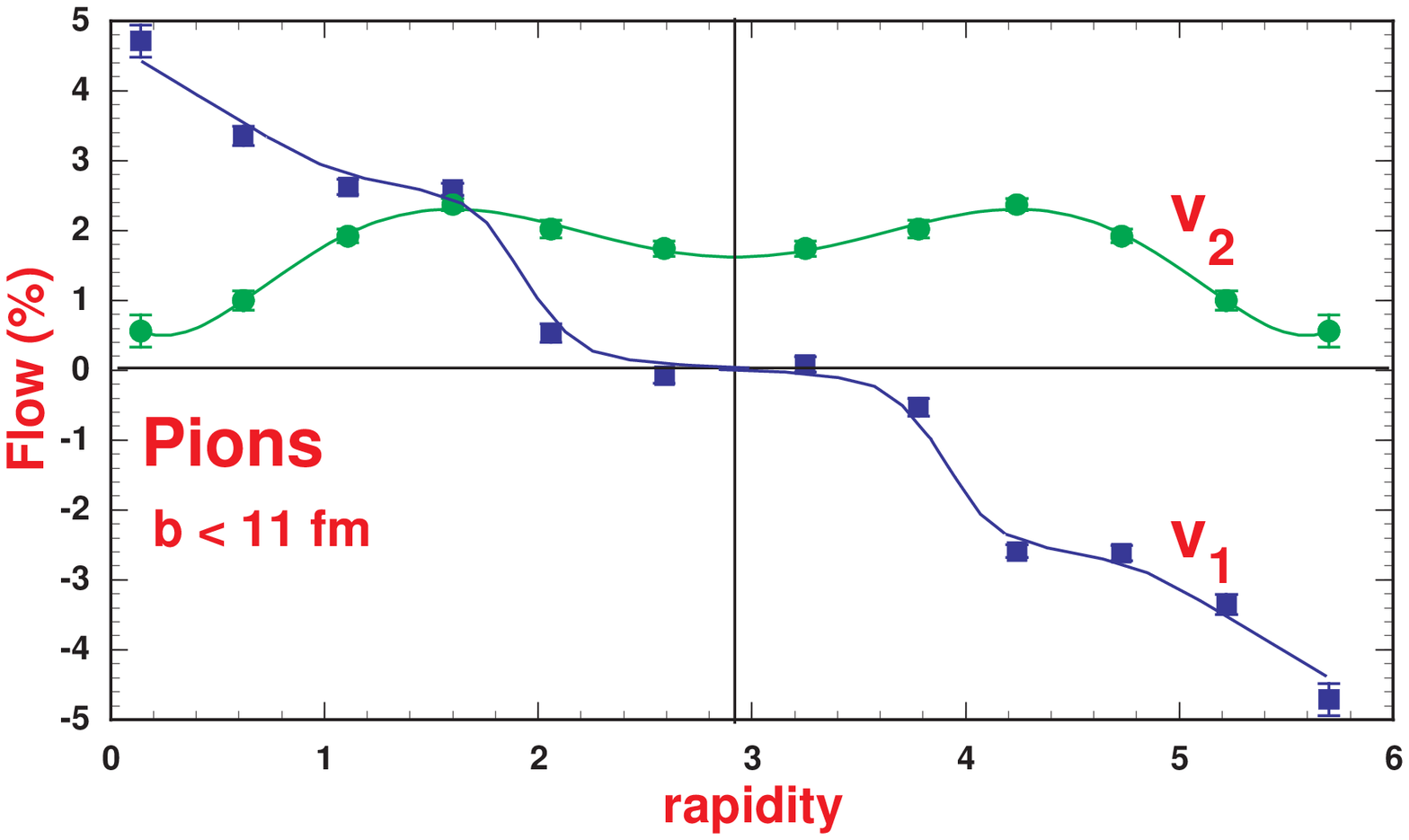,width=7.4cm}
\psfig{figure=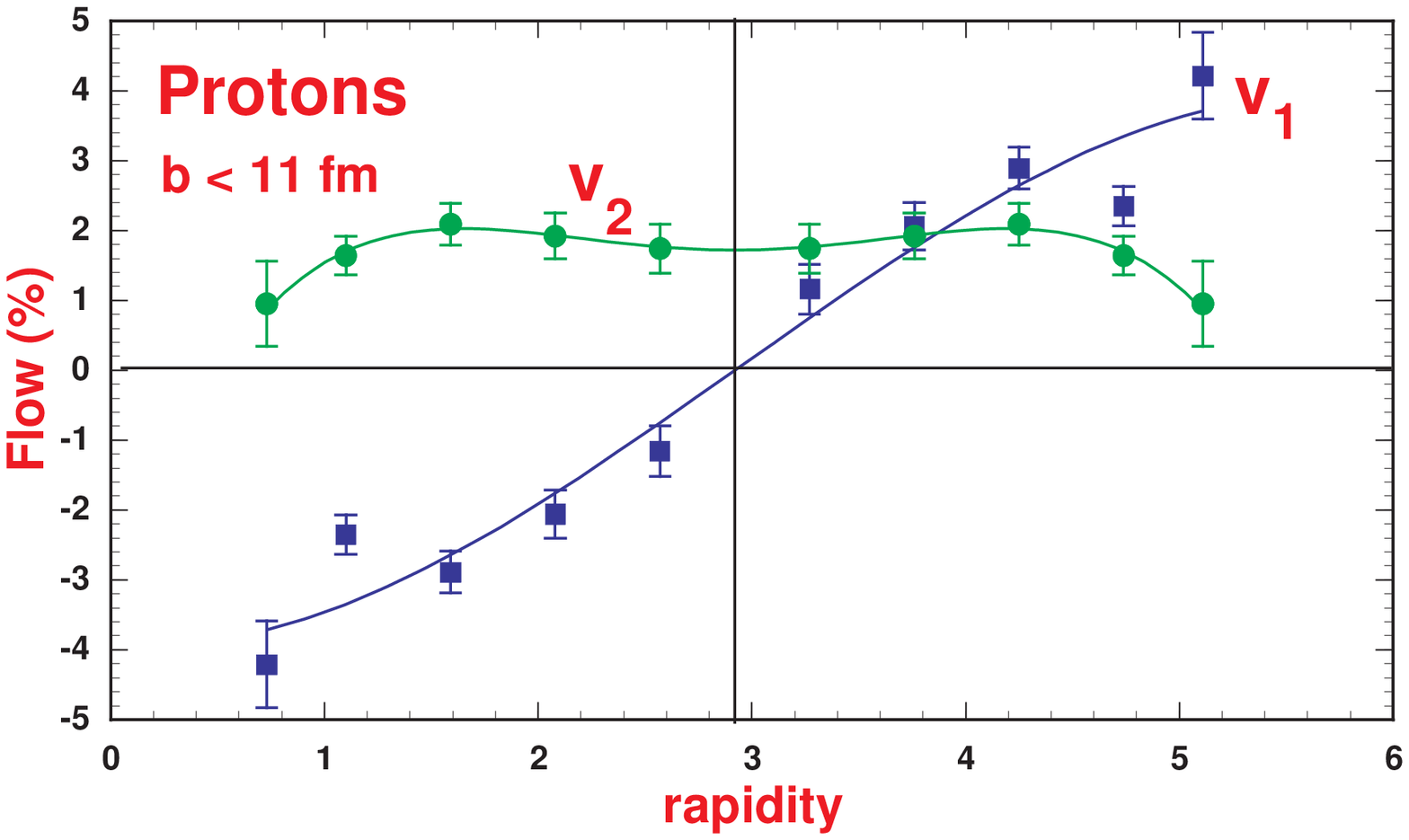,width=7.4cm}
\caption{Pion and proton directed and elliptic flow versus rapidity.}
\label{fig:v-y}
\end{minipage}
\hspace{\fill}
\begin{minipage}[t]{75mm}
\psfig{figure=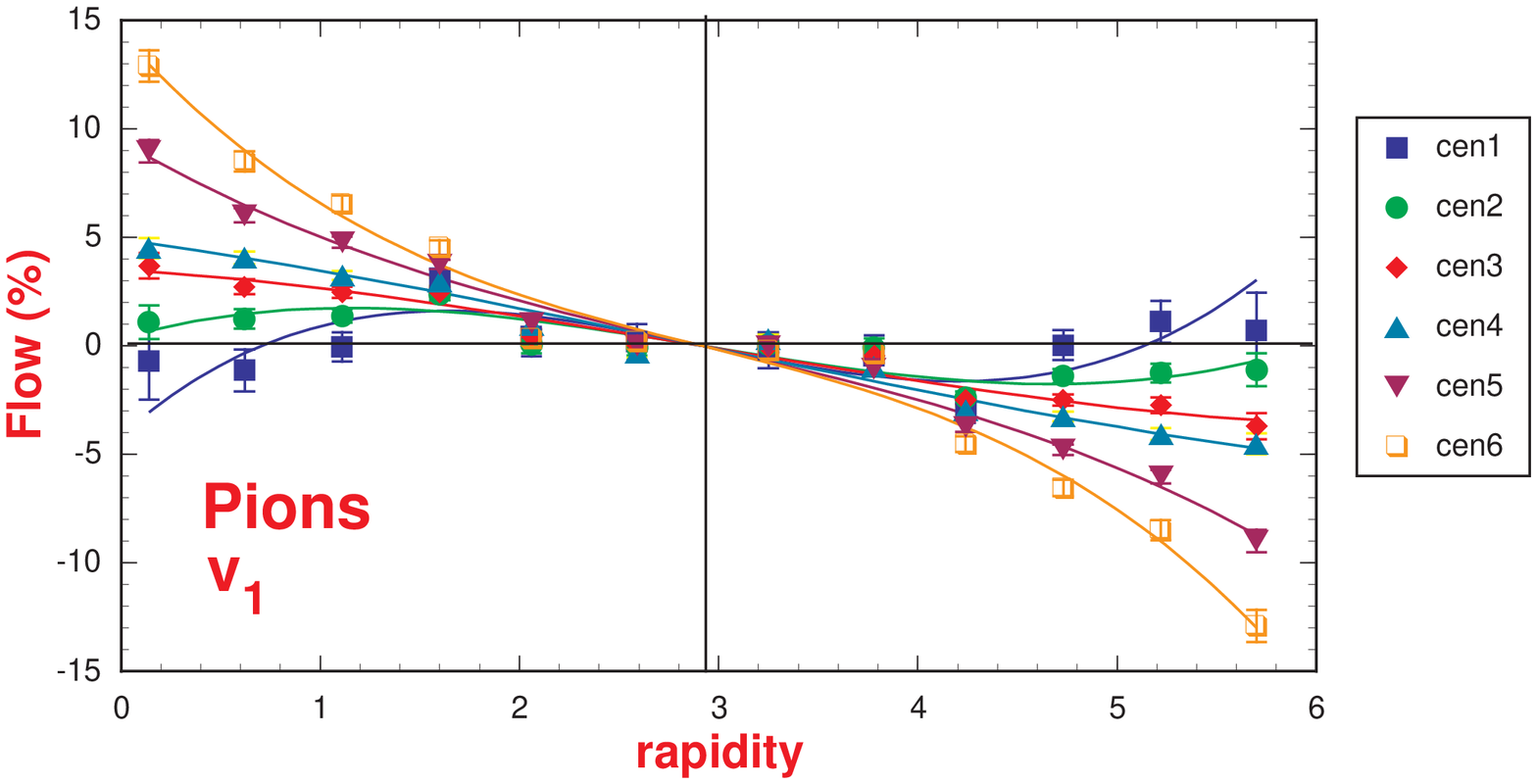,width=7.4cm}
\psfig{figure=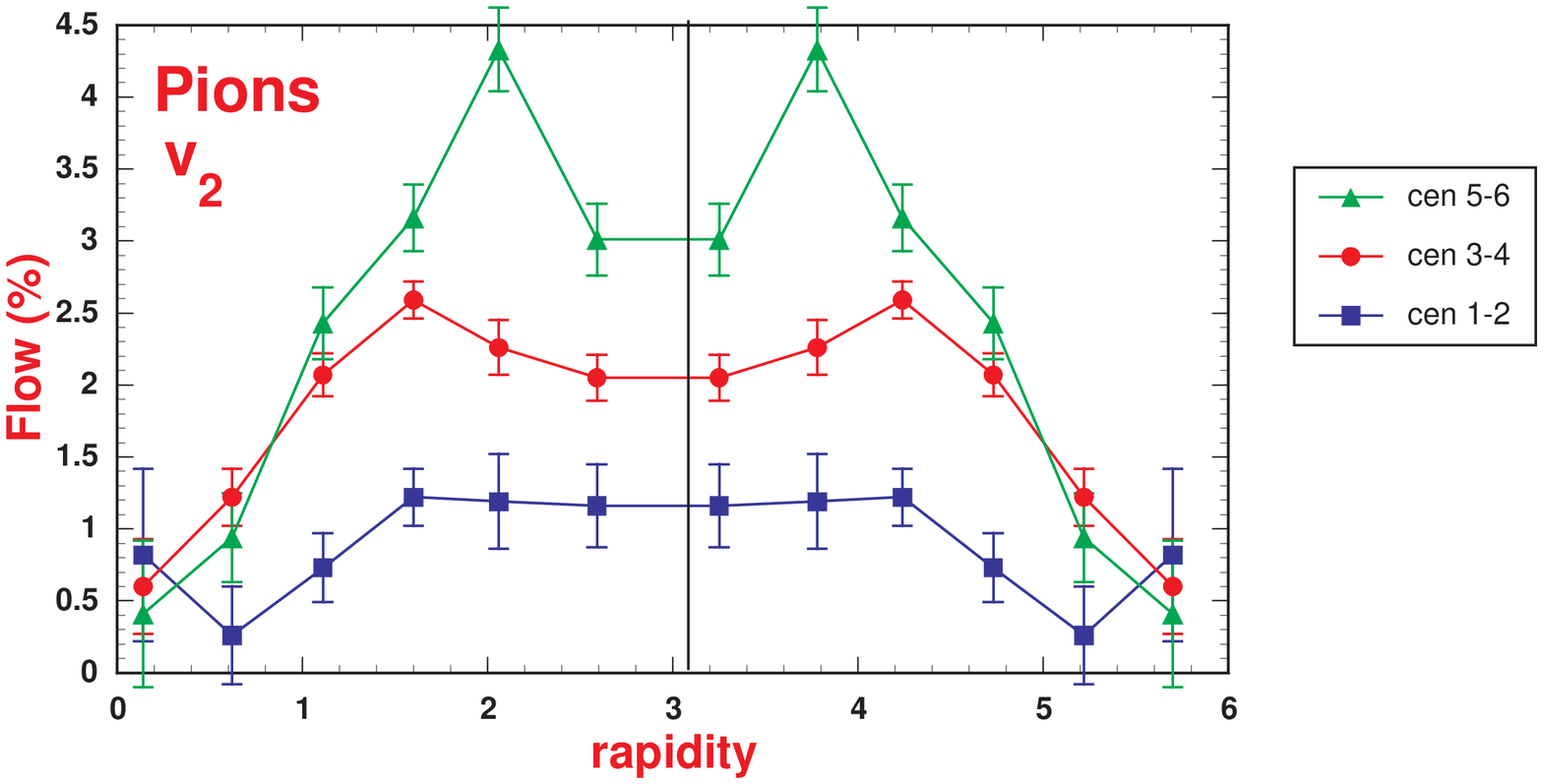,width=7.4cm}
\caption{Pion directed and elliptic flow for different centralities where
``cen1'' is the most central and ``cen6'' the most peripheral.}
\label{fig:cen}
\end{minipage}
\end{figure}

\begin{figure}[htb]
\begin{minipage}[t]{80mm}
\psfig{figure=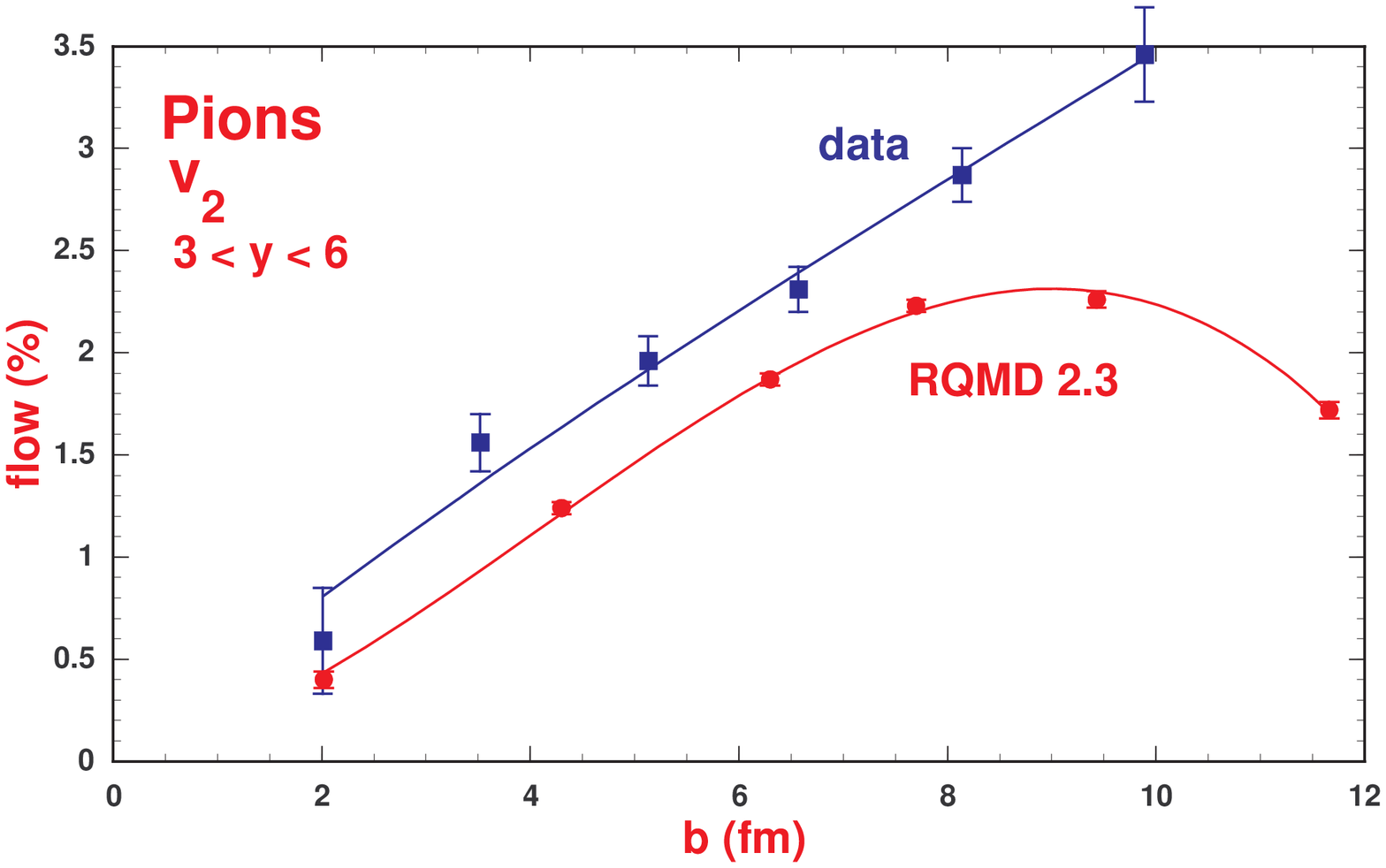,width=7.4cm}
\caption{Pion elliptic flow versus the impact parameter.} 
\label{fig:b}
\end{minipage}
\hspace{\fill}
\begin{minipage}[t]{75mm}
\psfig{figure=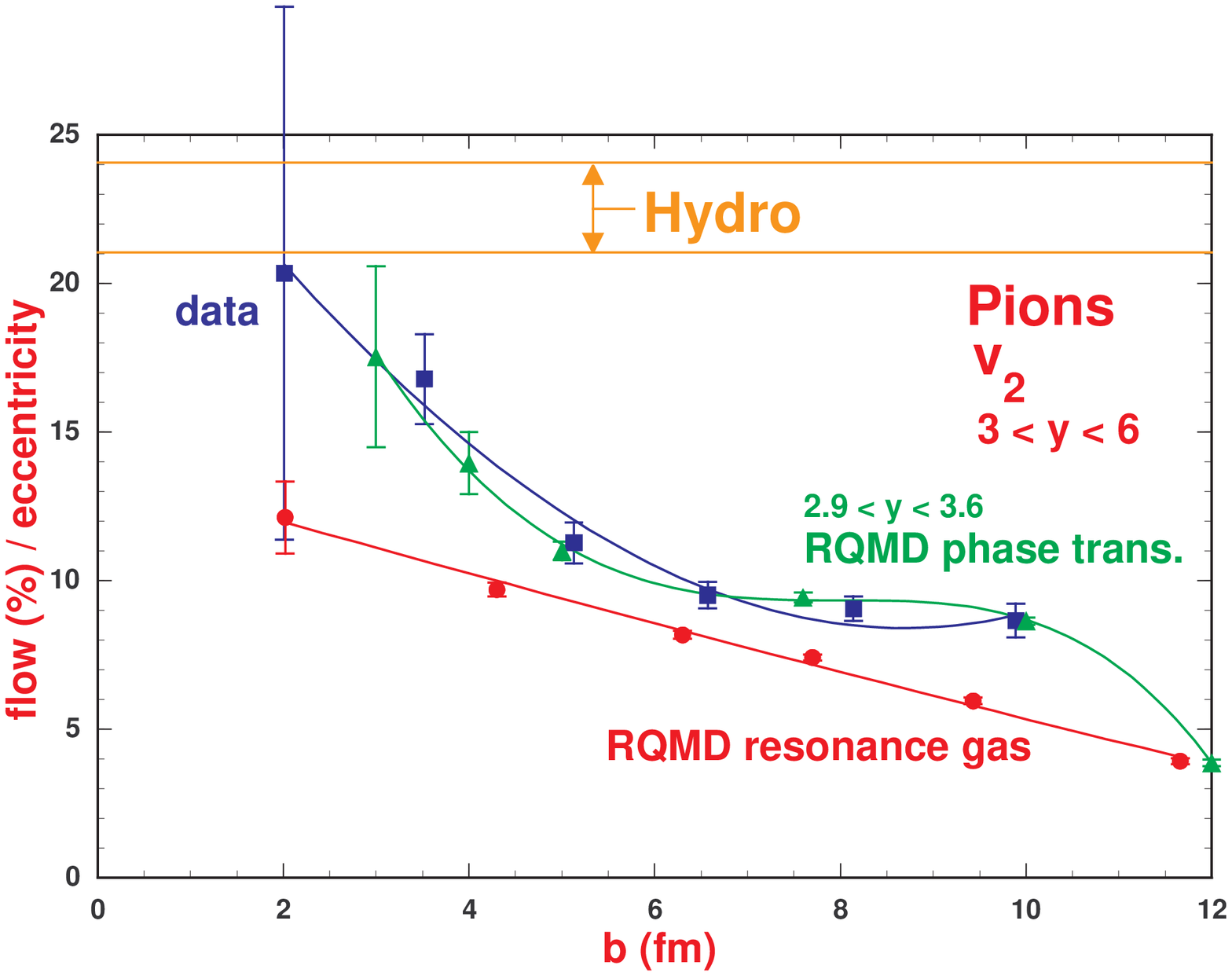,width=7.4cm}
\caption{Pion elliptic flow divided by the initial space elliptic anisotropy, 
$v_2/\eps$, versus the impact parameter.}
\label{fig:b-ep}
\end{minipage}
\end{figure}

The rapidity dependence of directed and elliptic flow integrated over
the whole range of measured impact parameters up to about 11 fm is
shown in Fig.~\ref{fig:v-y}. The pion $v_1$ values hug the axis near
mid-rapidity and the $v_2$ values for both pions and protons appear to
slightly peak somewhat away from mid-rapidity. For pions the $v_1$ and
$v_2$ values are shown for different centrality bins in
Fig.~\ref{fig:cen}. Both sets of flow values increase continuously as
the reaction becomes more peripheral. The elliptic flow values for
pions have been integrated over rapidity up to $y = 6$ and are shown
in Fig.~\ref{fig:b}, together with simulations from RQMD
v2.3~\cite{sorge2.3}. The flow from RQMD peaks at a medium impact
parameter whereas the flow from experiment continues to rise. In
Fig.~\ref{fig:b-ep} the $v_2$ values have been divided by the initial
space elliptic anisotropy~\cite{jacobs99}. In addition, results from
RQMD v3.0~\cite{sorge99} which includes a phase transition are
shown. Typical hydro results~\cite{kolb99} are also shown. The data
are below hydro indicating a lack of complete equilibration in the
reaction~\cite{voloshin99}. The data are above the RQMD resonance gas
and tantalizingly close to the RQMD phase transition
calculation. Clearly, it is important to process the full set of NA49
data and obtain final results.

\vspace{1.0cm}


\end{document}